\documentclass[final,5p,times,twocolumn]{elsarticle}
\usepackage{color}
\usepackage{graphicx}
\usepackage{dcolumn}
\usepackage{bm}
\usepackage{hyperref}
\usepackage{natbib}
\usepackage{graphicx}
\usepackage{amsmath}
\usepackage{amsfonts}
\usepackage{amsthm}
\usepackage{float}
\usepackage{subcaption}
\usepackage[bottom]{footmisc}
\usepackage{comment}
\usepackage{color}
\usepackage{appendix}
\usepackage{amssymb}
\usepackage{geometry}
\usepackage[normalem]{ulem}
\usepackage{amsmath}
\usepackage{amssymb}
\usepackage{graphicx}
\usepackage{color}
\usepackage{hyperref}

\usepackage[mathscr,scaled=1.15]{urwchancal}
\renewcommand\sout{\bgroup \color{red} \ULdepth=-.5ex \ULset}
\biboptions{sort&compress}
\journal{Physics Letters B}

\begin{document}

\begin{frontmatter}
\title{Principal Component Analysis of collective flow in Relativistic Heavy-Ion Collisions}

\date{today}

\author[ad1]{Ziming Liu}
\author[ad1,ad2]{Wenbin Zhao}
\author[ad1,ad2,ad3]{Huichao Song}
\ead{Huichaosong@pku.edu.cn}

\address[ad1]{Department of Physics and State Key Laboratory of Nuclear Physics and Technology, Peking University, Beijing 100871,China}
\address[ad2]{Collaborative Innovation Center of Quantum Matter, Beijing 100871, China}
\address[ad3]{Center for High Energy Physics, Peking University, Beijing 100871, China}

\begin{abstract}
In this paper, we implement Principal Component Analysis (PCA) to study the single particle distributions  generated from thousands of {\tt VISH2+1} hydrodynamic simulations with an aim to explore if a machine could directly discover flow from the huge amount of data without explicit instructions from human-beings. We found that the obtained PCA eigenvectors are similar to but not identical with the traditional Fourier bases. Correspondingly, the PCA defined flow harmonics $v_n^\prime$ are also similar to the traditional $v_n$  for $n=2$ and 3, but largely deviated from the Fourier ones for $n\geq 4$. A further study on the symmetric cumulants and the Pearson coefficients indicates that mode-coupling effects are reduced for these flow harmonics defined by PCA.
\end{abstract}
\end{frontmatter}


\section{\label{sec:level1}{Introduction}}

Collective flow is one of the most important observables in relativistic heavy-ion collisions, which provides valuable information on the initial state fluctuations, final state correlations and the QGP properties. In the past decades, various flow observables have been extensively measured in experiments and studied in theory~\cite{Teaney:2009qa,Romatschke:2009im,Huovinen:2013wma,Heinz:2013th,Gale:2013da,Song:2017wtw}. In general, these flow observables are defined based on the Fourier decomposition. For example, the integrated flow harmonics are defined as:
\begin{equation}
\label{fourier}
\begin{aligned}
\frac{{\rm d} N}{{\rm d} \varphi} &=\frac{1}{2\pi}  \sum_{-\infty}^{\infty}\vec{V}_n e^{-in\varphi}\\
&=\frac{1}{2\pi} (1+ 2 \sum_{n=1}^{\infty} v_{n} e^{-in(\varphi-\Psi_n)})
\end{aligned}
\end{equation}
where  $\vec{V}_n =v_ne^{in\Psi_n}$ is the $n$-th order flow-vector,  $v_{n}$ is the $n$-th order flow harmonics and $\Psi_{n}$ is the corresponding event plane angle. In general, the first coefficient, $v_1$, is called the {\emph{directed flow}},  the second coefficient, $v_2$, is called the {\emph{elliptic flow}} and the third coefficient $v_3$, is called  the {\emph{triangular flow}}. For $n \geq 3$, $v_n$ is also referred as the higher order flow harmonics.

In spite of the success of the flow measurements and the hydrodynamic descriptions, one essential question is why the Fourier expansion is a natural way to analyze the flow data. In this paper, we will address these questions with one of the machine learning techniques, called the Principal Component Analysis (PCA). In more details, we will investigate if a machine could directly discover flow from the huge amount of data of the relativistic fluid systems without explicit instructions from human beings.

PCA is one of the unsupervised algorithms of machine learning~\cite{DBLP:journals/corr/Shlens14} based on the Singular Value Decomposition (SVD) that diagonalize a random matrix with two orthogonal matrices. Compared with other deep learning algorithms, the advantage of PCA lies in its simple and elegant mathematical formulation, which is understandable and traceable to human beings, and is able to reveal the main structure of data in a quite transparent way.

Due to its strong power in data mining, PCA has been implemented to various research area of
physics~\cite{PhysRevLett.120.016601,lloyd2014quantum,PhysRevLett.111.083001,
PhysRevB.96.144432,PhysRevB.96.195138,Bhalerao:2014mua}.
In molecular dynamics, PCA has been utilized to distinguish break junction trajectories of single molecules~\cite{PhysRevLett.120.016601}, which is time efficient and can transfer to a wide range of multivariate data sets. In the field of quantum mechanics, the quantum version of PCA was applied to study quantum coherence among different copies
of the system~\cite{lloyd2014quantum}, which are exponentially faster
than any existing algorithm. In condensed matter physics, PCA has been implemented to study the phase transition in Ising model~\cite{PhysRevB.96.144432}, which found that eigenvectors of PCA can aid in the definition of the order parameter, as well as provide reasonable predictions for the critical temperature without any prior knowledge.  Besides, PCA is a widely used tool in engineering for model reduction to make computations more efficient~\cite{7394136}.

In relativistic heavy-ion collisions, PCA has been implemented to study the event-by-event flow fluctuations, using the 2-particle correlations with the Fourier expansion~\cite{Bhalerao:2014mua,Mazeliauskas:2015vea,Mazeliauskas:2015efa,Bozek:2017thv,Sirunyan:2017gyb}. Compared with the traditional method,  PCA explores all the information contained in the 2-particle correlations, which reveals the substructures in flow fluctuations~\cite{Bhalerao:2014mua,Mazeliauskas:2015vea,Mazeliauskas:2015efa}. It was found that the leading components of PCA correspond to the traditional flow harmonics and the sub-leading components evaluate the breakdown of the flow factorization at different $p_t$ or $\eta$ bins. Besides, PCA has also been used to study the non-linear mode coupling between different flow harmonics\cite{Bozek:2017thv}, which helps to discover some hidden mode-mixing patterns.
Recently, the CMS Collaboration further implemented PCA to analyze 2-particle correlation in Pb-Pb collisions at $\sqrt{s_{NN}}=$ 2.76 TeV and p-Pb collisions at $\sqrt{s_{NN}}=$ 5.02 TeV~\cite{Sirunyan:2017gyb}, showing the potential of largely implementing such machine learning technique to realistic data in relativistic heavy ion collisions.

These early PCA investigations on flow~\cite{Bhalerao:2014mua,Mazeliauskas:2015vea,
Mazeliauskas:2015efa,Bozek:2017thv,Sirunyan:2017gyb} are all based on the preprocessed
data with the Fourier expansion, which still belong to the category of traditional flow analysis.
In this paper, we will directly apply PCA to study the single particle distributions
from  hydrodynamic simulations without any priori Fourier transformation. We aim to explore
if PCA could discover flow with its own bases.

This paper is organized as follows. Sec.~II introduces relativistic hydrodynamics, principal component analysis (PCA)  and the corresponding flow analysis. Sec.~III shows and discusses the flow results from PCA and compares them with the ones from traditional Fourier expansion.  Sec.~IV summarizes and concludes the paper.

\section{Model and method}\label{Approach}
\subsection{{\tt VISH2+1 hydrodynamics}}
In this paper, we implement {\tt VISH2+1}~\cite{Song:2007fn,Song:2007ux,Song:2009gc,Shen:2014vra} to generate the final particle distributions for the PCA analysis. {\tt VISH2+1} \cite{Song:2007fn,Song:2007ux,Song:2009gc,Shen:2014vra} is a 2+1-dimensional viscous hydrodynamic code to simulate the expansion of the QGP fireballs, which solves the transport equations for the energy-momentum tenor $T^{\mu\nu}$ and the second order Israel-Stewart equations for the shear stress tensor $\pi^{\mu\nu}$ and bulk pressure $\Pi$ with an equation of state s95-PCE\cite{Bazavov:2014pvz,Bernhard:2016tnd} as an input. The initial profiles for {\tt VISH2+1} are provided by {{\tt {\tt TRENTo}}}, a parameterized initial condition model that generates event-by-event fluctuating entropy profiles with several tunable parameters~\cite{Moreland:2014oya,Bernhard:2016tnd}. These parameters, together with
the temperature dependent specific shear viscosity and bulk viscosity,  hydrodynamic starting time ($\tau_0=0.6 \ \mathrm{fm/c}$) and decoupling /switching temperature ($T_{sw} =148 \ \mathrm{MeV}$) have been fixed through fitting all charged and identified particle yields, the mean transverse momenta and the integrated flow harmonics in 2.76 A TeV Pb+Pb collisions using the Bayesian statistics\cite{Bernhard:2016tnd}, which also nicely described various flow data at the LHC~\cite{Zhao:2017yhj}. In practice, the transition from the hydrodynamic fluid to the emitted  hadrons on the freeze-out surface is realized by a Monte-Carlo event generator {\tt iss} based on the Cooper-Fryer formula\cite{Song:2010aq}:
\begin{equation}
\frac{dN}{dy p_T dp_T d\varphi} = \int_\Sigma \frac{g}{(2\pi)^3} p^\mu d^3 \sigma_\mu f(x, p)
\label{iSS.eq3}
\end{equation}
where $f(x, p)$ is the distribution function of particles,  $g$ is the degeneracy factor, and $d^3\sigma_\mu$ is the volume element on the freeze-out hypersurface.

For the following PCA analysis, as well as for the traditional flow analysis in comparison, we run the event-by-event {\tt VISH2+1} simulations with 12000 fluctuating initial conditions generated from {\tt {\tt TRENTo}} for 2.76 A TeV Pb-Pb collisions at 0\%-10\%,10\%-20\%, 20\%-30\%, 30\%-40\%, 40\%-50\% and 50\%-60\% centrality bins.  The default {\tt iss} sampling for each {\tt VISH2+1} simulation is 1000 events, which corresponds to the main results presented in Sec.~III.
In the appendix of this paper, we also investigate the ability of PCA to distinguish  signal and noise. We thus implement  25, 100 and 500 {\tt iss} samplings for each {\tt VISH2+1} simulation for such investigation. Note that the default  1000 {\tt iss} sampling used in this paper has already dramatically suppressed the statistical fluctuations from noises for the final hadron distributions. \\[-0.10in]

With the final particle distributions obtained from hydrodynamic simulations, various flow observables can be calculated based on the traditional flow harmonics defined by the Fourier decomposition in Eq.(\ref{fourier}). In Sec.III, the traditional flow  results will be served as the comparison to
the PCA results.

\begin{figure*}[t]
	\centering
	\begin{subfigure}[b]{0.60\textwidth}
		\centering
		\includegraphics[width=1.0\linewidth,height=5cm]{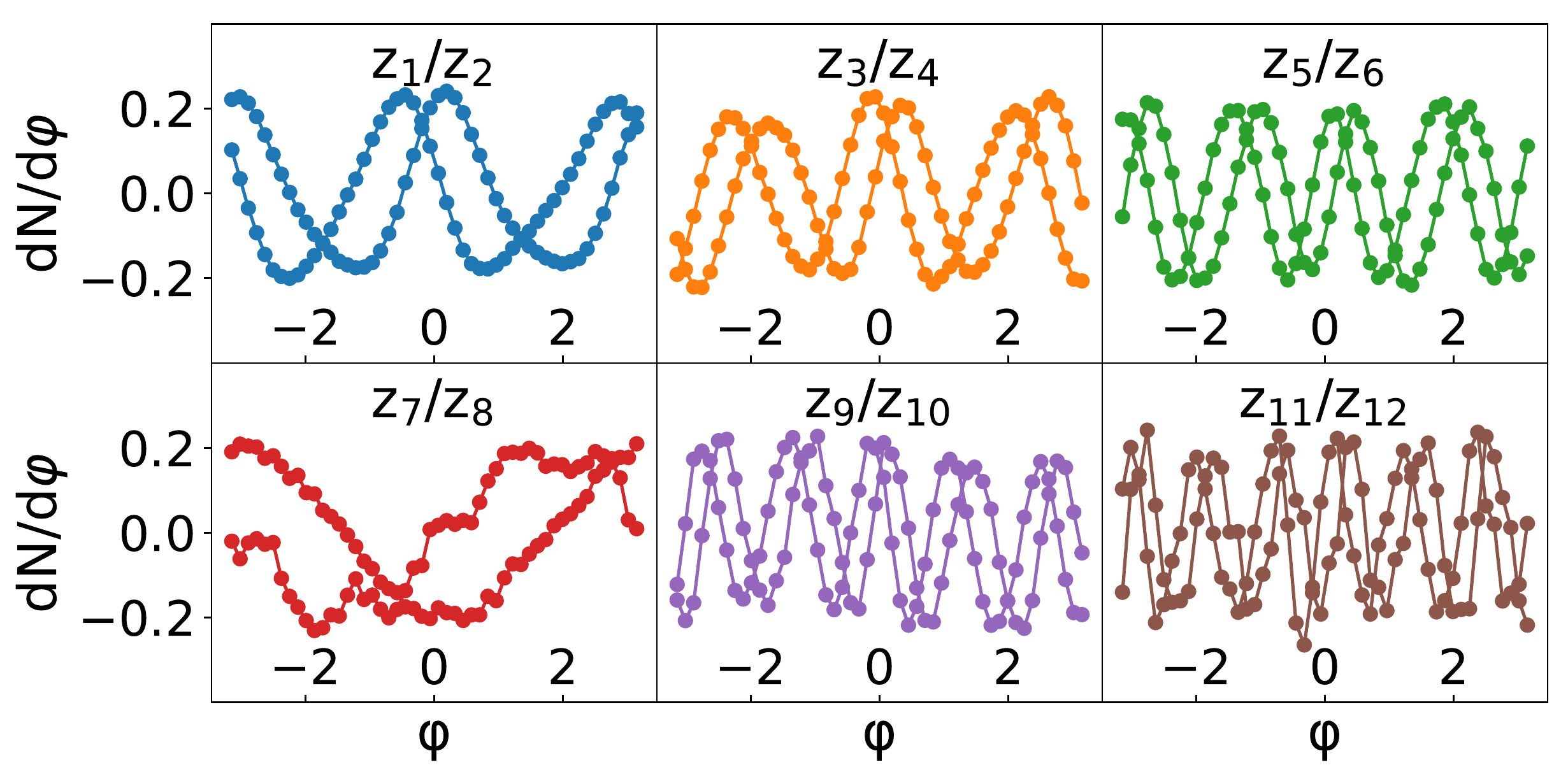}
		\caption{}
	\end{subfigure}%
	~
	\begin{subfigure}[b]{0.4\textwidth}
		\centering
		\includegraphics[width=1.0\linewidth,height=5.0cm]{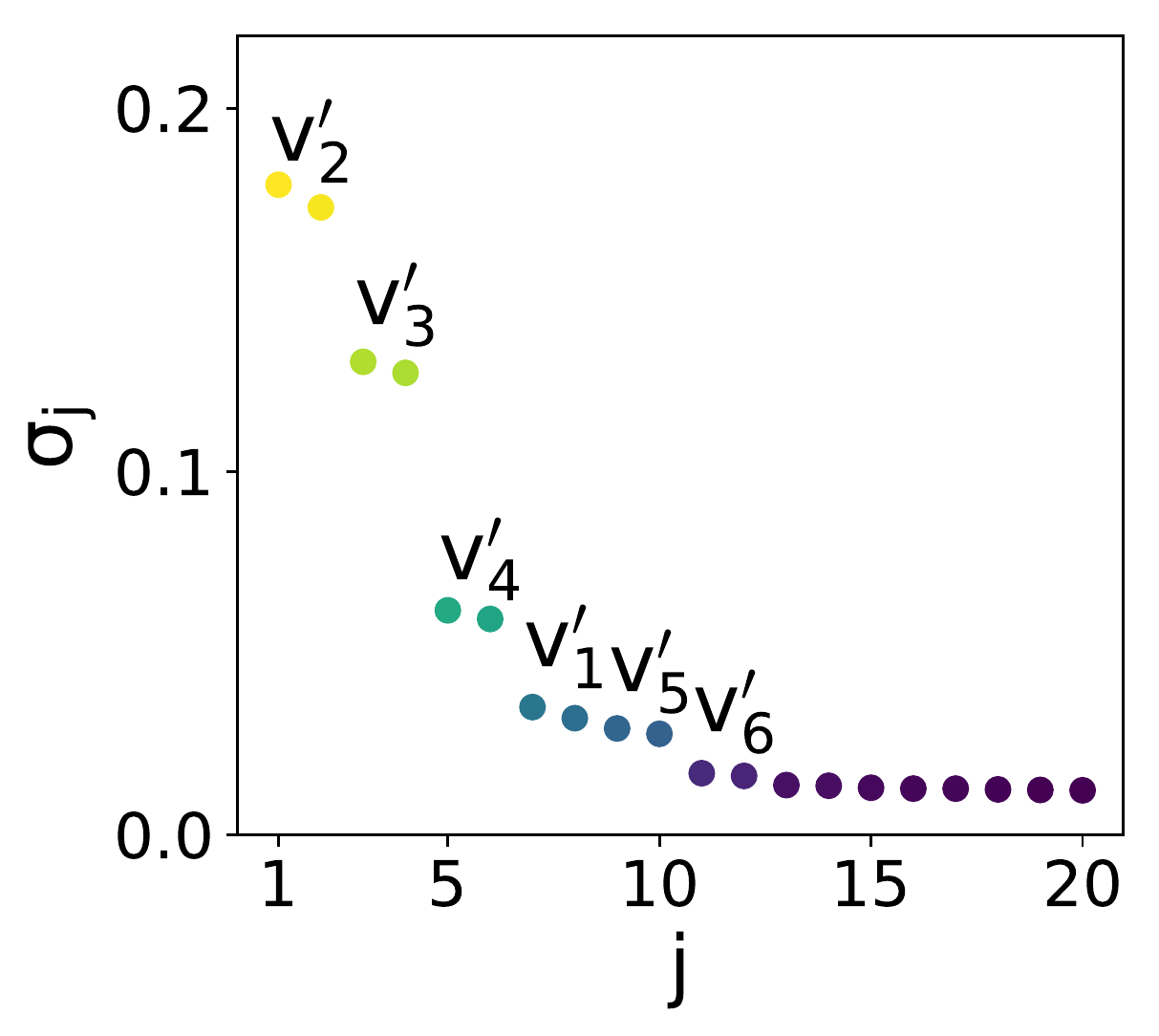}
		\caption{}
	\end{subfigure}
	\caption{(a) The first 12  eigenvectors ${z}_j \ (j=1,2,\cdots,12)$  and (b) the first 20 singular values $\sigma_j \ (j=1,2, ...,20)$, after applying PCA to the final state matrix $\mathbf{M_f}$.  The matrix $\mathbf{M_f}$ is constructed from 2000 $dN/d\varphi$ distributions, generated from the event-by-event {\tt VISH2+1} simulations with {\tt {\tt TRENTo}} initial conditions for 10\%-20\% Pb+Pb collisions at $\sqrt{s_{NN}}=$ 2.76 A TeV.}
	\label{fig:eigenmodes}
\end{figure*}

\subsection{{Principal Component Analysis (PCA)}}

Principal Component Analysis (PCA) is a statistical method to analyze complicated data, which aims to transform a set of correlated variables into various independent variables via orthogonal transformations. These obtained main eigenvectors, associated with large or unnegligible singular values, are also called the principal components, which reveal the most representative characteristics of the data.  In practice, PCA implements the Singular Value Decomposition (SVD) to a real matrix, which obtains a diagonal matrix with the diagonal elements arranged in a descending order. Therefore, one needs to first construct a related matrix before the following PCA and SVD analysis. Since this paper
focuses on investigating the integrated flow with PCA, such final state matrix $\mathbf{M_f}$ is constructed
from the angular distribution of all charged hadrons $dN/d\varphi $ $(|y|<1.0)$ (obtained from Eq.(\ref{iSS.eq3})) of $N=2000$ independent events in each centrality bin,  using {\tt VISH2+1} simulations with {\tt {\tt TRENTo}} initial conditions. In more details, we divide the azimuthal angle $[-\pi,\pi]$ into $m=50$ bins and count the number of particles in each bin.  For the $j_{th}$ bin in event ($i$),  the  number of particles is denoted as $dN/d\phi^{(i)}_j $, which is also the $i_{th}$ row and $j_{th}$ column of the matrix $\mathbf{M_{f}}$~\footnote{In practice, we normalize the event vector in $\mathbf{M_{f}}$  to get rid of the multiplicity fluctuations.}.

Then, we apply SVD to the  final state matrix $\mathbf{M_f}$ with the size $N\times m$ (Here, $N=2000$ and $m=50$), which gives
\begin{eqnarray}
\quad\quad\quad\quad \mathbf{M_{f}}=\mathbf{{X}{\Sigma}{Z}}=\mathbf{{V}{Z}}
\end{eqnarray}
where $\mathbf{{X}}$ and $\mathbf{{Z}}$ are two orthogonal matrices with the size of $N\times N$ and $m\times m$, respectively. $\mathbf{{\Sigma}}$ is a diagonal matrix with diagonal elements (singular values) arranged in the descending order $\sigma_1>\sigma_2>\sigma_3 \ \cdots >0$.

With such matrix multiplication, the $i_{th}$ row of matrix $\mathbf{M_f}$, denoted as $dN/d\varphi^{(i)}$, can be expressed by the linear combination of the eigenvectors $z_j$ (the $j_{th}$ row of matrix $\mathbf{Z}$) with $j=1,2,... ,m $:
\begin{eqnarray}
dN/d\varphi^{(i)}&=&\sum_{j=1}^m {x}_j^{(i)}{\sigma}_j {z}_j
=\sum_{j=1}^m \tilde{v}_j^{(i)} {z}_j  \nonumber  \\
&\approx & \sum_{j=1}^{{k}} \tilde{v}_j^{(i)} {z}_j \ \ \ (i)=1,... ,N
\label{pca}
\end{eqnarray}
where $(i)=1,2,... , N$, represents the index of the event, $m$ is the number of angular bins of the inputting events. $\tilde{v}_j^{(i)}$ is the corresponding coefficient of ${z}_j$ for the $i_{th}$ event. In the spirit of PCA, one only focuses on the most important components, so there is a cut at the indices ${k}$ in the last approximation of Eq.(\ref{pca}).
In Sec.~III, we will show that ${k}=12$ is a proper truncation for the integrated flow analysis, and the shape of the bases or eigenvectors ${z}_j \ (j=1, ...,{k})$ is similar to but not identical with the Fourier transformation bases $\cos(n\varphi)$ and $\sin(n\varphi)$ ($n=1,... ,6$) used in the traditional method. Correspondingly,
$\tilde{v}_j^{(i)} \ (j=1,..., {k})$  is identified as the real or imaginary part of the flow harmonics for event ($i$), and the singular values ${\sigma}_j$ are associated with the corresponding event averaged flow harmonics at different orders. For more details, please also refer to Sec.~III.

%

\section{Results}\label{Results and Discussions}

In this section, we implement PCA to analyze the single particle distributions $dN/d\varphi$ from hydrodynamics simulations in Pb+Pb collisions at $\sqrt{s_{NN}}=$ 2.76 A TeV. Firstly, we focus on
the singular values, eigenvectors as well as the associated
coefficients of PCA and explore if such unsupervised learning could discover flow with
its own bases.

\begin{table}[t]
	\centering
		\caption{Event averaged flow harmonics $v_n'$ from PCA and $v_n$ from the Fourier expansion,
		for {\tt VISH2+1} simulated Pb+Pb collisions at 10-20\% centrality.}
	\begin{tabular}{|c|c|c|c|}
		\hline
		$n$ & $\overline{v_n^\prime}${(PCA)} & $\overline{v_n^\prime}\times 10^2$ & $\overline{v_n}\times 10^2$ \\ [0.5ex]
		\hline
		2 & $\sqrt{\frac{m}{2}}\sqrt{\sigma_1^2+\sigma_2^2}$ & 6.03 & 6.08 \\\hline
		3 & $\sqrt{\frac{m}{2}}\sqrt{\sigma_3^2+\sigma_4^2}$ & 2.57 & 2.53 \\\hline
		4 & $\sqrt{\frac{m}{2}}\sqrt{\sigma_5^2+\sigma_6^2}$ & 1.21 & 1.25 \\\hline
		5 & $\sqrt{\frac{m}{2}}\sqrt{\sigma_9^2+\sigma_{10}^2}$ & 0.57 & 0.66 \\\hline
		6 & $\sqrt{\frac{m}{2}}\sqrt{\sigma_{11}^2+\sigma_{12}^2}$ & 0.26 & 0.37 \\
		\hline
	\end{tabular}
	\label{table:comp relation}
\end{table}

In practice, we run 2000 event-by-event {\tt VISH2+1} hydrodynamic simulations with {\tt {\tt TRENTo}} initial conditions to generate the $dN/d\varphi$ distributions for 10\%-20\% Pb+Pb collisions at $\sqrt{s_{NN}}=$ 2.76 A TeV.
With these $dN/d\varphi$ distributions, we construct the final state matrix $\mathbf{M_f}$ and then
implement SVD to  $\mathbf{M_f}$ as described in Sec.~II.   Fig.~\ref{fig:eigenmodes}  shows these obtained first 12 eigenvectors ${z}_j \ (j=1,2, ...,12)$ and the first 20 singular values ${\sigma}_j \ (j=1,2, ...,20)$ of PCA, arranged by the descending order of magnitude~\footnote{Each eigenvector is automatically normalized with
	$||z_j||_2^2=\sum_{i=1}^m (z_j)_i^2=1$ ($m=50$), due to the orthogonality of the eigenvector matrix $\mathbf{Z}$.}. As introduced in Sec.~II, these eigenvectors contain the most representative information on correlations among final particles. Fig.~\ref{fig:eigenmodes} shows that the $1{st}$ and $2{nd}$ eigenvectors from PCA are similar to the Fourier decomposition bases $\mathrm{sin}(2\varphi)$ and $\mathrm{cos}(2\varphi)$, and the $3{rd}$ and $4{th}$ components are similar to $\mathrm{sin}(3\varphi)$ and $\mathrm{cos}(3\varphi)$, etc.  Meanwhile, Fig.~\ref{fig:eigenmodes} (b) shows that singular values
${\sigma}_j \ (j=1,2, ...,12)$ are arranged in pairs. These results indicate that each pair of the singular values may associate with the real and imaginary parts of the event averaged flow vectors at different orders. Therefore,
we define the event averaged flow harmonics of PCA with these paired singular values, as outlined in the
the second column of Table \ref{table:comp relation}. The values of these PCA flow at different order
are compared with the traditional flow harmonics from the Fourier expansion in Table \ref{table:comp relation},
which are close, but not exactly the same values for $n\leq6$.


\begin{figure*}[t]
	\includegraphics[width=0.95\linewidth]{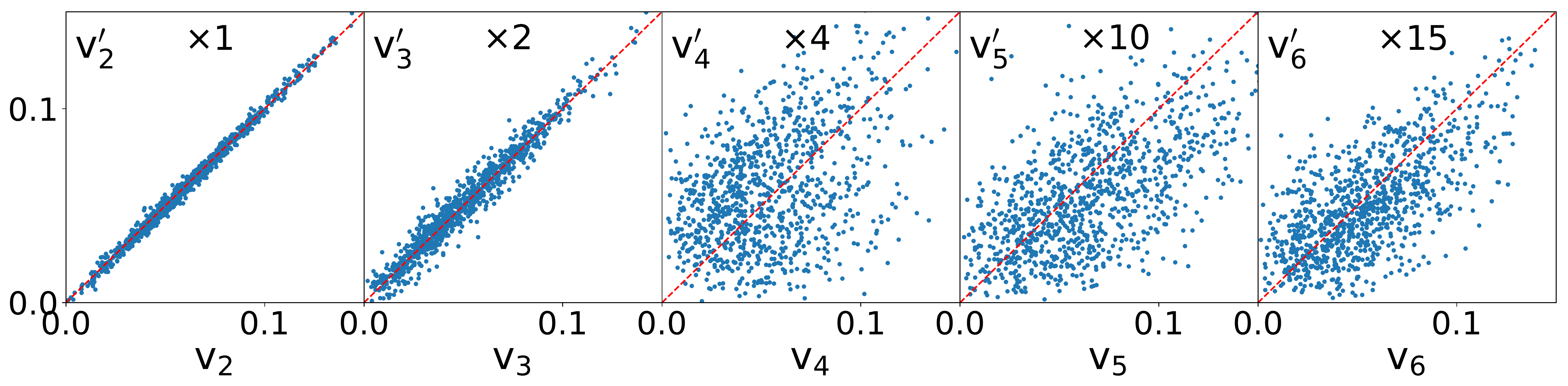} 
	\caption{A comparison between the  event-by-event flow harmonics $v_n'$ from PCA and $v_n$ from the Fourier expansion, for {\tt VISH2+1} simulated  Pb+Pb collisions at 10-20\% centrality.}
	\label{fig:vv_compare}
\end{figure*}

\begin{figure*}[t]
	\centering
	\includegraphics[width=1.0\linewidth]{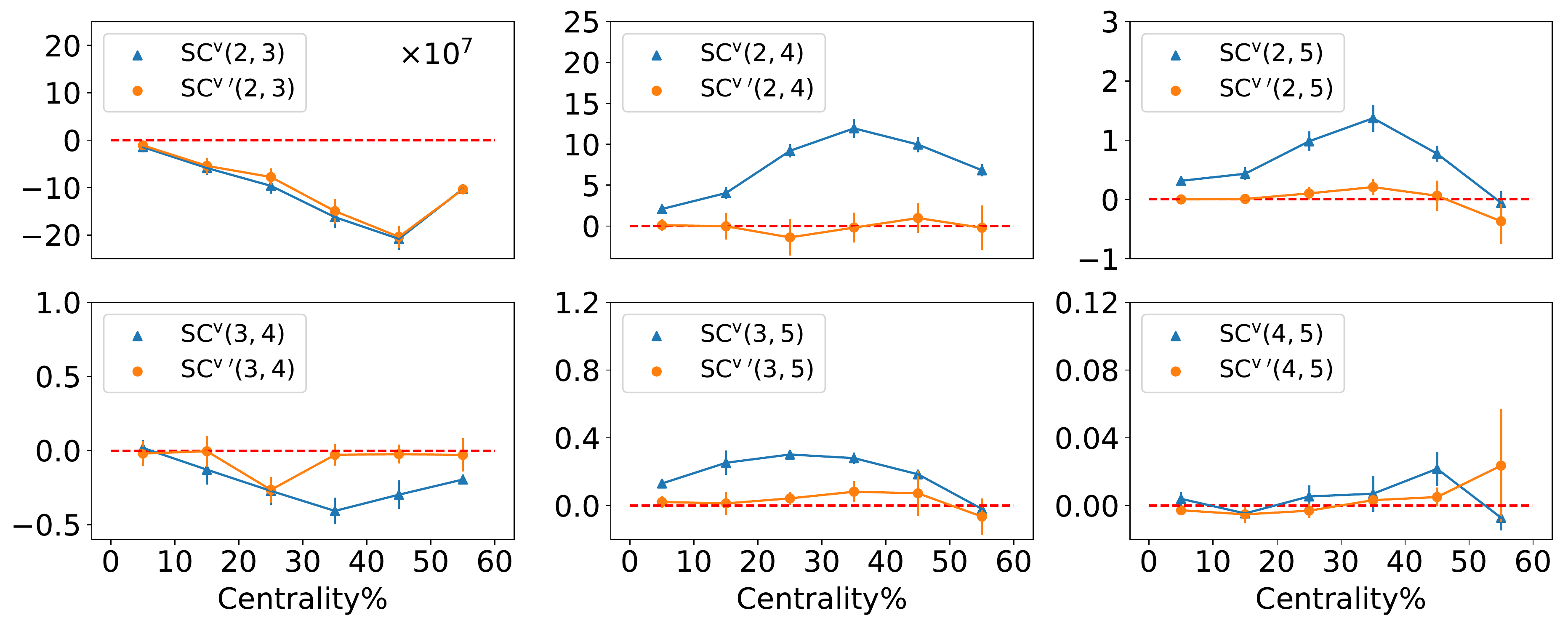}
	\caption{ Symmetric Cumulants $SC^v{'(m,n)}$ from PCA and $SC^v{'(m,n)}$ from the Fourier expansion, for {\tt VISH2+1} simulated Pb+Pb collisions at various centralities.}
	\label{fig:vv_correlation}
\end{figure*}

\begin{figure*}[t]
	\centering
	\includegraphics[width=1.0\linewidth]{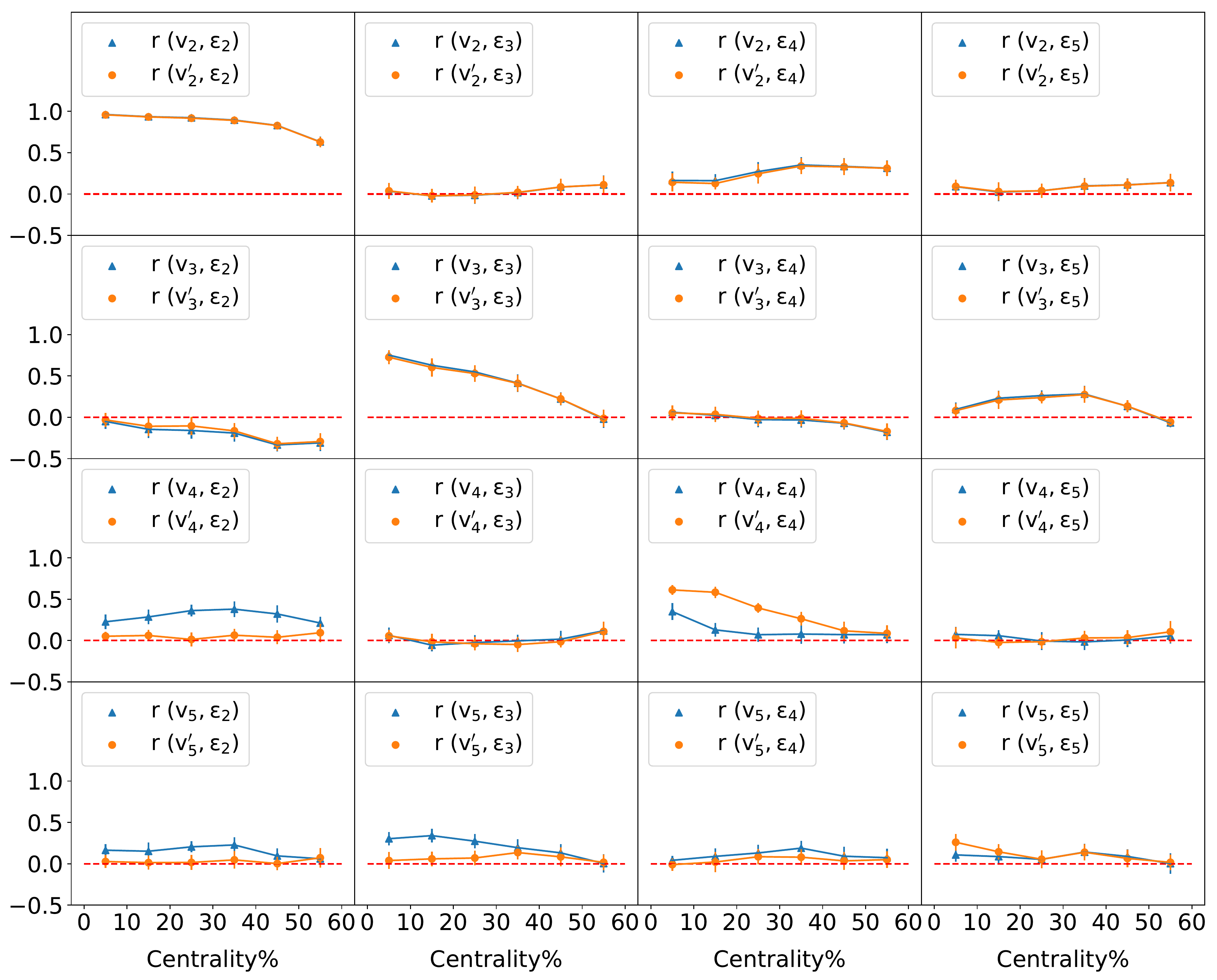}
	\caption{The Pearson coefficient $r(v'_n, \varepsilon'_m)$ from PCA and $r(v_n, \varepsilon_m)$
		from Fourier expansion, for {\tt VISH2+1} simulated Pb+Pb collisions at various centralities.}
	\label{fig:ve_correlation}
\end{figure*}

As explained in Sec II, one could also read the event-by-event flow harmonics from the results of PCA. In more details, such PCA flow harmonics for event ($i$) is associated with these coefficients $\tilde{v}_j^{(i)}, j=1...k$ in Eq.~(\ref{pca}). Therefore, we define the event-by-event flow harmonics $v_n^\prime$ with magnitudes projected onto PCA bases, similar to the event averaged ones defined in Table \ref{table:comp relation}. For example, $v_2^\prime=\sqrt{\frac{m}{2}}\sqrt{\tilde{v}_1^2+\tilde{v}_2^2}$ and $v_3^\prime=\sqrt{\frac{m}{2}}\sqrt{\tilde{v}_3^2+\tilde{v}_4^2}$ ($m=50$), etc. Fig.~2 compares $v_n'$ from PCA and $v_n$ from  the traditional Fourier expansion at different orders. For the event-by-event elliptic flow $v_2$ and $v_2'$ and triangular flow $v_3$ and $v_3'$, the definitions from PCA and that from Fourier expansion are highly agree with each other, which mostly fall on the diagonal lines. For these higher order flow harmonics with $n\geq4$, these PCA results are largely deviated from the traditional Fourier ones. We also noticed that the first two PCA eigenvector  ${z}_1$ and ${z}_2$ for $v_2'$ are similar to but not identical with the Fourier bases $\mathrm{sin}(2\varphi)$ and $\mathrm{cos}(2\varphi)$ with $n=2$, which contain the contributions from $\mathrm{sin}(4\varphi)$ and $\mathrm{cos}(4\varphi)$. Similarly, the PCA eigenvectors  ${z}_3$ and ${z}_4$ also contain the contributions from other Fourier bases.   Such mode mixing in the PCA eigenvectors leads to the large deviations between $v_4$ and $v_4'$, as well as between $v_5$ and $v_5'$, etc.

To evaluate the correlations between different PCA flow harmonics $v_m'$ and $v_n'$, we  calculate the symmetric cumulants as once defined for traditional flow harmonics~\cite{ALICE:2016kpq,Bhalerao:2014xra,Zhu:2016puf}:
\begin{eqnarray}
SC^v{'(m,n)} &=&\left<v_m^{\prime2} v_{n}^{\prime2}\right>-\left<v_{m}^{\prime2}\right>\left<v_{n}^{\prime2}\right>.
\label{eq:4p_sc_cumulant}
\end{eqnarray}
Correspondingly, the traditional symmetric cumulants $SC^v{(m,n)}$ just replace $v'_m$ and $v'_n$ with $v_m$ and $v_n$ from the Fourier expansion.

FIG.~\ref{fig:vv_correlation} compares the symmetric cumulants $SC^v{'(m,n)}$ from PCA and $SC^v{'(m,n)}$ from Fourier
expansion, for the event-by-event {\tt VISH2+1} simulations in 2.76 A TeV Pb+Pb collisions at various centrality bins.
One finds that, except for $SC^v(2,3)$, almost all PCA symmetric cumulants $SC^v{'(m,n)}$  reduce significantly compared to the traditional ones. Although $v'_4$ from PCA largely deviated from the traditional $v_4$ from the Fourier expansion, the obtained $SC^v{'(2,4)}$ shows a significant suppression, which contradicts to the long believed idea that the nonlinear hydrodynamics evolution strongly couples $v_2^2$ to $v_4$, leading to an obvious positive correlations between $v_2$ and $v_4$ obtained from Fourier expansion.  Similarly, the non-linear mode coupling  between $v'_2$ and $v'_5$, $v'_3$ and $v'_5$ and $v'_3$ and $v'_4$ for these PCA defined flow harmonics also decrease, which results in the reduced symmetric cumulants $SC^v{'(2,5)}$, $SC^v{'(3,5)}$ and  $SC^v{'(3,4)}$ correspondingly.

To evaluate the correlations between the initial and final state fluctuations,  we use the Pearson coefficients $r(v'_n, \varepsilon_m)$ and $r(v_n, \varepsilon_m)$ to characterize the linearity between the PCA flow harmonics $v'_n$ and the initial eccentricities $\varepsilon_m$,  as defined as the following:
\begin{equation}
r(v'_n, \varepsilon_m)=\frac{\langle v'_n \varepsilon_m \rangle -\langle v'_n \rangle \langle \varepsilon_m \rangle}{\sqrt{(v'_n-\langle v'_n\rangle)^2(\varepsilon_m-\langle \varepsilon_m\rangle)^2}}
\end{equation}
Here, $\varepsilon_m$ is the traditional eccentricities defined by Eq.(\ref{eq:eccCalculation}). In Appendix A,
we will demonstrate that, with a properly chosen smoothing procedure, the
event-by-event eccentricities $\varepsilon'_m$ from PCA is highly similar
to $\varepsilon_m$ from the traditional method. We thus use $\varepsilon_m$
in the Pearson coefficient definition  $r(v'_n, \varepsilon_m)$ for PCA. Meanwhile, we can
also calculate the Pearson coefficient $r(v_n, \varepsilon_m)$ for the
traditional flow with Fourier expansion, which just replaces the flow
harmonics $v'_n$ in Eq.~(6) by $v_n$. According to the definition, the Pearson
coefficient falls in the range $[-1,1]$, with $r>0$ implying a positive correlation,
and $r<0$ implying a negative correlation.

Fig.~\ref{fig:ve_correlation} plots the Pearson coefficients $r(v'_n, \varepsilon_m)$ from PCA and $r(v_n, \varepsilon_m)$ from the Fourier expansion, for {\tt VISH2+1} simulated Pb+Pb collisions at various centralities. With these Pearson coefficients, we focus on evaluating if the PCA defined flow harmonics reduce or increase
the correlations with the corresponding initial eccentricities.
As shown in Fig.~\ref{fig:vv_correlation}, the event-by-event  flow harmonics $v'_2$ or $v'_3$ from PCA are approximately equal to the Fourier ones  $v_2$ or $v_3$. As a result, these Pearson coefficients involved with these two flow harmonics  $r(v'_2, \varepsilon_m)$ and $r(v'_3, \varepsilon_m)$ are almost overlap with the Fourier ones $r(v_2, \varepsilon_m)$ and $r(v_3, \varepsilon_m)$ as shown by these upper panels in the first two rows.  Meanwhile,  these diagonal Pearson coefficients $r(v'_2, \varepsilon_2)$ or $r(v_2, \varepsilon_2)$ and $r(v'_3, \varepsilon_3)$ or $r(v_3, \varepsilon_3)$ are much larger than other ones, which confirms the early conclusion that the elliptic flow and triangular flow are mainly influenced by the initial eccentricity $\varepsilon_2$ and $\varepsilon_3$ with the approximate linear relationship $v_2 \thicksim \varepsilon_2$ ($v'_2 \thicksim \varepsilon_2$) and $v_3 \thicksim \varepsilon_3$ ($v'_3 \thicksim \varepsilon_3$)~\cite{Qiu:2011iv,Teaney:2012ke}.

Although $v'_4$ from PCA is largely deviated from the traditional $v_4$ in Fig.~\ref{fig:vv_correlation}, such PCA definition largely enhances  correlations between  $\varepsilon_4$, and also largely reduces the correlations between $\varepsilon_2$.  For example, at  20-30\% centrality, the Pearson coefficients $r(v_4, \varepsilon_4)$ is only 70\% of the $r(v_4^\prime, \varepsilon_4)$, while $r(v_4, \varepsilon_2)$ is 200\% larger than $r(v'_4, \varepsilon_2)$. Traditionally, it is generally believed that $v_4$ is largely influenced by  $\varepsilon_2^2$  through the non-linear evolution of hydrodynamics. Our PCA analysis showed that such mode mixing could be deduced through a redefined PCA bases. Meanwhile, such PCA defined bases also significantly reduce the mode mixing for other higher order flow harmonics such as  between $v'_5$ and $\varepsilon_2$, $v'_5$ and $\varepsilon_3$, etc.

\section{Conclusions}\label{sec:5}
In this paper, we implemented Principal Components Analysis (PCA) to study the single particle distributions of thousands of events generated from {\tt VISH2+1} hydrodynamic simulations. Compared with the early PCA
investigations on flow that imposed the Fourier transformation in the input data~\cite{Bhalerao:2014mua,Mazeliauskas:2015vea,Mazeliauskas:2015efa,Bozek:2017thv,Sirunyan:2017gyb}, we focused on analyzing the raw data of hydrodynamics and exploring if a machine could directly discover flow from the huge amount of data without explicit instructions from human-beings. We found that the PCA eigenvectors are similar to but not identical with the traditional Fourier basis. Correspondingly, the obtained flow harmonics $v_n^\prime$ from PCA are also similar to the traditional $v_n$ for $n=2$ and 3, but largely deviate from the Fourier ones for $n\geq 4$. With these PCA flow harmonics, we found that, except for $SC^v{'(2,3)}$, almost all other symmetric cumulants $SC^v{'(m,n)}$ from PCA decrease significantly compared to the traditional $SC^v{(m,n)}$. Meanwhile, some certain Pearson coefficients $r(v'_n, \varepsilon_m)$ that evaluate the linearity between the PCA flow harmonics and the initial eccentricities are obviously enhanced (especially for $n \geq 4$), together with an corresponding reduction of the off-diagonal elements.

These results indicate that PCA has the ability to discover flow with its own basis, which also reduce the related mode coupling effects, when compared with traditional flow analysis based on the Fourier expansion. We emphasis that these  eigenvectors from PCA are modeled to be orthogonal and uncorrelated to each other. As a result, most of the symmetric cumulants $SC^v{'(m,n)}$ from PCA that evaluate the correlations between different flow harmonics are naturally reduced compared with the traditional ones. Besides, the PCA flow harmonics $v'_n$ presents an enhanced linear relationship to the corresponding eccentricities $\varepsilon_n$, especially for $n=4$. These results seem contradictory to the long believed idea that hydrodynamics evolution are highly non-linear, which leads to strong mode-coupling between different flow harmonics. Our PCA investigation has shown that such mode coupling effects could be reduced with new-defined bases for the flow analysis. With such finding, the non-linearity of the relativistic
hydrodynamic systems created in heavy ion collisions should be re-evaluated, which we would like to further explore it with such PCA method in the near future.

\section*{Acknowledgements}
We would like to thank the fruitful discussions with J.~Jia, R.~Lacey, D.~Teaney and M.~Zhou . This work is supported by the NSFC and the MOST under grant Nos. 11675004, 11435001 and 2015CB856900. We also gratefully acknowledge the extensive computing resources provided by the Super-computing Center of Chinese Academy of Science (SCCAS), Tianhe-1A from the National Supercomputing Center in Tianjin, China and the High-performance Computing Platform of Peking University.

\begin{appendices}
	\appendix
	\section{{PCA} for initial profiles with smoothing procedure }
In this appendix, we focus on analyzing the initial state fluctuations using the PCA method. Traditionally, the initial state fluctuations are evaluated by the eccentricity coefficients $\varepsilon_{n}$ at different order, which are defined as~\cite{Qiu:2011iv}:
	\begin{equation}
	\label{eq:eccCalculation}
	\varepsilon_{n}e^{in\Phi_{n}}=-\frac{\int r\, dr\, d\varphi\, r^n\, e^{in\varphi}\, s(r, \varphi)}{\int r\, dr\, d\varphi\, r^n\, s(r, \varphi)},
	\end{equation}
where $\Phi_{n}$ is the participant plan angle,  $s(r,\varphi)$ is the initial entropy density and $\varphi$ is the azimuthal angle in the transverse plane~\cite{Qiu:2011iv}.
	
For the PCA analysis, we first construct the initial state matrix $\mathbf{M_i}$, using the azimuthal angle distribution of the initial entropy $dS/d\varphi$ which is defined by
	\begin{equation}
	\frac{dS}{d\varphi}=\int r^2drs(r,\varphi)
	\end{equation}
obtained from 2000 event-by-event {\tt TRENTo} initial conditions.  A direct PCA analysis shows that more than 100 eigenvectors are needed to capture the rich structures of the initial state fluctuations. In contrast, 12 PCA eigenvectors are enough to describe the final state ones since the hydrodynamic evolution tends to smear out inhomogeneity of the evolving systems. In order to connect and compare these PCA singular values from the initial and final states,  we implement a smoothing procedure for the initial profiles before the PCA analysis.
	
\begin{figure*}[t]
		\includegraphics[width=0.95\linewidth]{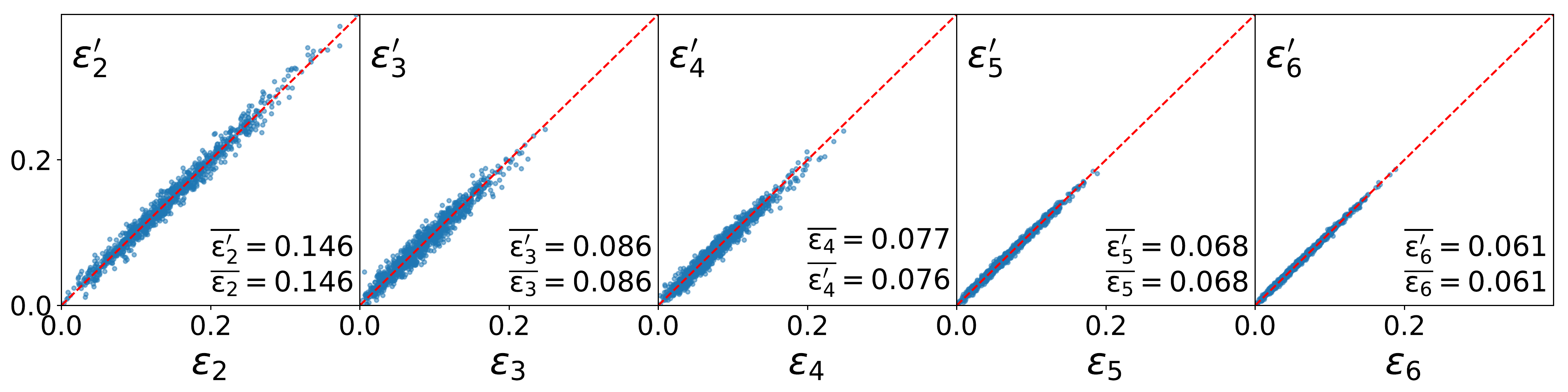}
		\caption{A comparison between the  event-by-event eccentricites $\varepsilon'_n$ from PCA and $\varepsilon_n$ from the traditional definition (A1), for {\tt TRENTo} initial conditions at 10-20\% centrality. In the corresponding panels, we also write the values of event averaged eccentricites $\overline{\varepsilon'_n}$ from PCA and $\overline{\varepsilon_n}$ from the traditional definition. }
		\label{fig:ee_compare}
\end{figure*}	
		
In more details, we apply a circular convolution with  to the initial density profile $dS/d\varphi$, which is written as:
	\begin{equation}
	(\frac{dS}{d\varphi})_{smooth}=\int_{-\pi}^{\pi} K(\varphi^\prime,\varphi)\frac{dS}{d\varphi^\prime}d\varphi^\prime
	\end{equation}
Here, $K(\varphi^\prime,\varphi)$ is the convolution kernel, which is taken a gaussian form $K(\varphi^\prime,\varphi)=\frac{1}{\sqrt{2\pi}a}e^{-\frac{(\varphi^\prime-\varphi)^2}{2a^2}}$. Here, we fine tune the radius $a$ to ensure the the same decaying rate for the PCA singular values from the initial profiles and final profiles. The obtained optimized value for $a$ is 0.251 rad.
	
With such smoothing procedure, we reconstruct the initial state matrix $\mathbf{M_i}$ with 2000 event-by-event $(\frac{dS}{d\varphi})_{smooth}$ distributions from {\tt TRENTo} for each selected centralities. As the case for the flow analysis in Sec. II B and Sec.III,  the implementation of SVD and PCA to the initial state matrix, $\mathbf{M_i}=\mathbf{\hat{Y}\hat{\Sigma}\hat{Z}}=\mathbf{\hat{E}}\mathbf{\hat{Z}}$, gives the singular value $\hat{\sigma_j}$, eigenvectors $\hat{z_j}$ and the corresponding eccentricity coefficients $\hat{\varepsilon}_j^{(i)}$, $(j=1,...\hat{k})$ such that
	\begin{eqnarray}
	dS/d\varphi^{(i)}&=&\sum_{j=1}^m {y}_j^{(i)}{\hat{\sigma}}_j {\hat{z}}_j
	=\sum_{j=1}^m \hat{\varepsilon}_j^{(i)} {\hat{z}}_j  \nonumber  \\
	&\approx & \sum_{j=1}^{{k}} \hat{\varepsilon}_j^{(i)} {\hat{z}}_j  \ \ \qquad  (i)=1,... ,N
	\end{eqnarray}
We find that the PCA eigenvectors $\hat{z_j}$ of the initial states are highly similar to traditional Fourier bases $\cos(2\varphi)$, $\sin(2\varphi)$, $\cos(3\varphi)$, $\sin(3\varphi)$, etc. Meanwhile, we could associate the singular value  $\hat{\sigma}_j$ to the event averaged initial eccentricities of PCA, $\bar{\varepsilon}'_n$, at different orders
and connect the coefficients $\hat{\varepsilon}_j^{(i)}, (j=1,...\hat{k})$ to the real or imaginary part of the PCA event-by-event initial eccentricities $\varepsilon'_n (n=1,...\hat{k}/2)$ as the case for flow~\footnote{Here, $\bar{\varepsilon}'_2$=$\sqrt{\frac{m}{2}}\sqrt{\hat{\sigma}_3^2+\hat{\sigma}_4^2}$,  $\bar{\varepsilon}'_3$=$\sqrt{\frac{m}{2}}\sqrt{\hat{\sigma}_5^2+\hat{\sigma}_6^2}$,  $\bar{\varepsilon}'_4$=$\sqrt{\frac{m}{2}}\sqrt{\hat{\sigma}_7^2+\hat{\sigma}_8^2}$, etc.
with $m=50$ the number of bins. For event-by-event definition of ${\varepsilon}'_n (n=1,...\hat{k}/2)$, we could simply replace $\hat{\sigma}_j$ with $\hat{\varepsilon}_j$ $(j=1,...\hat{k})$, correspondingly.}.
		
Fig.~\ref{fig:ee_compare} compares the event-by-event eccentricites $\varepsilon'_n$ from PCA and $\varepsilon_n$ from the traditional definition (A.1) for the {\tt TRENTo} initial conditions at 10-20\% centrality. It shows, with a properly chosen smoothing procedure of the initial conditions, $\varepsilon'_n$  and  $\varepsilon_n$ agree with each other well till ${n=6}$ . Meanwhile, the event averaged eccentricites  $\overline{\varepsilon'_n}$ from PCA and $\overline{\varepsilon_n}$ from (\ref{eq:eccCalculation}) also fit each other very well, which is much better than ones for flow shown in Table \ref{table:comp relation} and Fig.~\ref{fig:vv_compare}.  Therefore, for the investigation of initial state and final state correlations, we only use the traditional $\varepsilon_m$ to define the  Pearson coefficient $r(v'_n, \varepsilon_m)$, and $r(v_n, \varepsilon_m)$ for both PCA and traditional flow in Sec.III.

\section{Signal and noise distinguishment from PCA}
In the event-by-event {\tt VISH2+1} simulations, both initial state fluctuations
and statistical fluctuations from the {\tt iss} particle sampling during the Cooper-Fryer
freeze-out influence the  emissions and distributions of final particles. It is generally believed
that the hydrodynamic evolution translate the initial state fluctuations into final
state correlations, which directly relate to flow signals. {Meanwhile, the statistical
fluctuations during Cooper-Fryer freeze-out with a finite number of particle emission introduce
statistical noise for the flow definition in each event.} As a result, flow harmonics from
traditional Fourier expansion are generally analyzed with an event average of millions of events.
For the event-by-event flow analysis, one implements the standard Bayesian unfolding procedure  to suppress effects from the finite multiplicites and non-flow \cite{Aad:2013xma}.
	\begin{figure}
		\centering
		\includegraphics[width=0.75\linewidth,height=5cm]{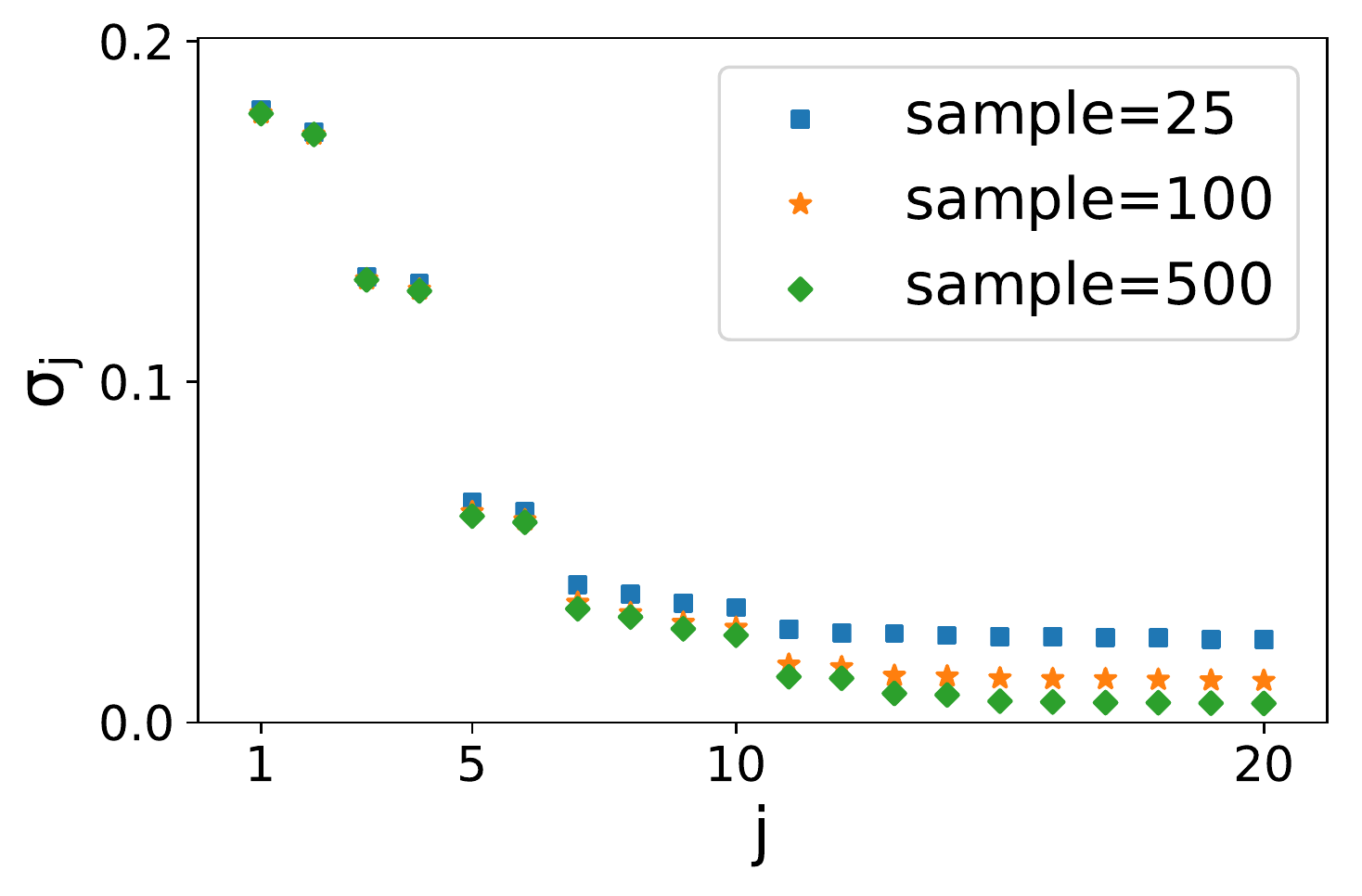}
		\caption{The first 20 singular values of PCA ${\sigma}_j \ (j=1,2,\cdots,20)$ for the final state matrixes  $\mathbf{M_f}$ with different weighted signal and noise. Such matrix $\mathbf{M_f}$ are constructed with
			the $dN/d\varphi$ distributions from 2000 event-by-event {\tt VISH2+1} simulations with 25, 100 and 500 {\tt iss} samplings.}
		\label{fig:tail}
	\end{figure}

In this appendix, we further explore the ability of PCA to distinguish the signal and noise. With such purpose,
we implement 25, 100 and 500 {\tt iss} samplings for each {\tt VISH2+1} simulation to generate the
$dN/d\varphi$ distributions of final particles and the related final state matrixes $\mathbf{M_f}$ with different weighted signal and noise. Then, we implement PCA to analyze these matrixes.  As shown in Fig.~\ref{fig:tail}, the distribution of the PCA singular values is changed with the number of  {\tt iss} samplings. For these systems with large statistical fluctuations, for example with 25 {\tt iss} samplings, the singular values $\sigma_j$ at large $j$ tend to have a long and high tail. For these systems with reduced statistical fluctuations with more {\tt iss} samplings, the height of the tail is largely decreased. Meanwhile, we noticed that these eigenvectors with an index $j$ smaller than a certain ``magic number"(12 in this case) is signal-like which has a basis similar to the Fourier one, while these eigenvectors with larger $j$ behave so randomly and chaotically, that we associate these eigenvectors with the noise patterns of the systems.Besides, we check the height of these PCA tails  and found the ratios among these heights for different iss samplings approximately satisfy  $\frac{1}{\sqrt{25}}:\frac{1}{\sqrt{100}}:\frac{1}{\sqrt{500}}$, such relation is known as the Law of Large Numbers
for statistical noise. With more number of samplings, the height of  the tail
would further decrease. In the main part of this paper, we thus set the {\tt iss} samplings for each {\tt VISH2+1}
simulation to 1000, which largely suppresses the noise effects from the statistical fluctuations
and makes PCA analysis focus on studying flow signal itself.

\end{appendices}

\section*{References}

\bibliography{ref}

\end{document}